\documentclass[journal]{IEEEtran}
\usepackage{graphicx}

\ifCLASSINFOpdf
\else
\fi

\begin{document}

\title{Combined 3D PET and Optical Projection Tomography Techniques for Plant Root Phenotyping}

\author{Qiang~Wang,~\IEEEmembership{Member,~IEEE,}
        Sergey~Komarov,
        Aswin~J.~Mathews,~\IEEEmembership{Student member,~IEEE,}
        Ke~Li,~\IEEEmembership{Student member,~IEEE,}
        Christopher~Topp,
        Joseph~A.~O'Sullivan,~\IEEEmembership{Fellow,~IEEE,}
        and~Yuan-Chuan~Tai,~\IEEEmembership{Member,~IEEE}
\thanks{This work is supported in part by the DOE-BER (DE-SC0005157), NSF (DBI-1040498), National Academies Keck Futures Initiative, WU I-CARES, and MIR internal funds. The PET image reconstruction is conducted by the WU Center for High Performance Computing that are funded in part by NIH (RR031625, RR022984)
}
        
\thanks{Q. Wang, S. Komarov and Y.-C. Tai are with the Department
of Radiology, Washington University in St. Louis, St. Louis,
MO, 63110 USA.}
\thanks{A. Mathews, K. Li and J. A. O'Sullivan are with the Department of
 Electrical and Systems Engineering, Washington University in St. Louis,  MO, 63130 USA.}
\thanks{C. Topp is with the Donald Danforth Plant Science Center, St. Louis, MO, 63132 USA.}}

\maketitle

\begin{abstract}
New imaging techniques are in great demand for investigating underground plant roots systems which play an important role in crop production. Compared with other non-destructive imaging modalities, PET can image plant roots in natural soil and produce dynamic 3D functional images which reveal the temporal dynamics of plant-environment interactions. In this study, we combined PET with optical projection tomography (OPT) to evaluate its potential for plant root phenotyping. We used a dedicated high resolution plant PET imager that has a 14 cm diameter transaxial and 10 cm axial field of views, and multi-bed imaging capability. The image resolution is around 1.25 mm using ML-EM reconstruction algorithm. B73 inbred maize seeds were germinated and then grown in a sealed jar with transparent gel-based media. PET scanning started on the day when the first green leaf appeared, and was carried out once a day for 5 days. Each morning, around 10 mCi of $^{11}$CO$_2$ was administrated into a custom built plant labeling chamber. After 10 minutes, residual activity was flushed out with fresh air before a 2-h PET scan started. For the OPT imaging, the jar was placed inside an acrylic cubic container filled with water, illuminated with a uniform surface light source, and imaged by a DSLR camera from 72 angles to acquire optical images for OPT reconstruction. The same plant was imaged 3 times a day by the OPT system (morning, noon and evening). Plant roots growth is measured from the optical images. Co-registered PET and optical images indicate that most of the hot spots appeared in later time points of the PET images correspond to the most actively growing root tips. The strong linear correlation between $^{11}$C allocation at root tips measured by PET and eventual root growth measured by OPT suggests that we can use PET as a phenotyping tool to measure how a plant makes subterranean carbon allocation decisions in different environmental scenarios.
\end{abstract}


\IEEEpeerreviewmaketitle

\section{Introduction}

\IEEEPARstart{W}{ith} the global climate change and increasing demand for food crops and renewable energy sources, developing crops that can improve or sustain yields  in harsh environments are becoming an integral part of the solution to this worldwide challenge. While modern sequencing technologies offer unprecedented information about genotype, the phenotypic outcomes of plant x environment interactions are hardly predictable. To close this knowledge gap, plant phenotyping is an increasingly important area of research. 
Many imaging based technologies have been adapted to measure multiple parameters of a plant under various conditions with high-throughput. For example, the Scanalyzer (LemnaTec GmbH) bench-top system can image hundreds of \textit{arabidopsis} grown in well plates each day, while a custom imaging system at the Danforth Center (Bellwether Foundation Phenotyping Facility, http://www.danforthcenter.org/scientists-research/core-technologies/phenotyping) can image a thousand plants in the growth chamber a day by automated conveyor system and imaging stations.

While imaging of leaves and structures above ground have been made easier by these technologies, studying of belowground roots remain a challenge. Some groups use transparent media such as gel to grow plants to visualize root structure. Most of the current root imaging systems are light-based\cite{Clark2013}\cite{Slovak2014} and can only provide morphological information of the subjects. Thus, imaging methods that reveal physiological information non-invasively with good temporal resolution are of great value to enrich the tool sets for future plant pheontyping\cite{Li2014}.

PET is a functional and molecular imaging technique that provides \textit{in vivo} measurement of dynamic radio-tracer distribution in a whole plant non-invasively. These dynamic PET images reveals the temporal physiological process happening inside plant. With plants grown in a transparent gel media, the anatomical change of plant root can be precisely captured by a low-cost optical imaging system\cite{Topp2013}. The study presented here explores the potential applications of this combined multi-modality imaging system on plant root phenotyping. 


\section{Materials and methods}

\subsection{Plant PET imaging system}
\begin{figure}[h]
\centering
\includegraphics[width=0.5\textwidth]{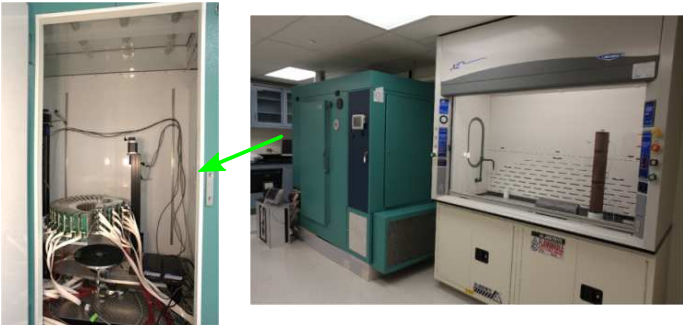}
\caption{ A dedicated plant PET imager (Left) seats inside a growth chamber (right). A fume hood adjacent to the plant growth chamber and lead lined radioactive gas delivery lines are used for radio-tracer administration.}
\label{fig:PETSystem}
\end{figure}

The dedicated plant PET system\cite{Wang2014} as shown in Figure \ref{fig:PETSystem} is designed with two unique features: (1) configurable system geometry to accommodate plants of different sizes and shapes; (2) the ability to control the environment in which the plants will be grown and studied. 

This plant PET system also provides $\sim$1.25 mm spatial resolution, which is especially important for imaging of small young plants with complex roots structure. The system sensitivity at center of field of view (FOV) is 1.3\%. The imager has 15 cm trans-axial and 10 cm axial FOV. With the automatic radio-active gas delivery system, the same subjects can be imaged repeatedly without disturbance which is important for plant studies that are very sensitive to environmental change.

\subsection{Optical projection tomography system}
\begin{figure}
\centering
\includegraphics[width=0.5\textwidth]{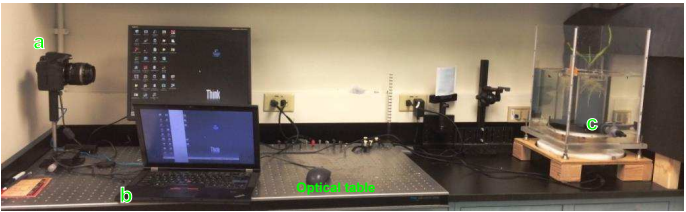}
\caption{ Optical imaging system: (a) DSLR camera; (b) controlling laptop for synchronizing object rotation and image capturing; (c) water tank and rotation stage for optical imaging.}
\label{fig:OPTSystem}
\end{figure}

Figure \ref{fig:OPTSystem} shows the setup of the optical projection tomography (OPT) system in our plant imaging lab. A maize plant is germinated and grown inside a glass cylindrical jar filled with transparent gel. During an optical imaging experiment, this jar is seat inside a rectangle water tank to compensate refraction induced distortion in optical images. A small rotation stage controls the rotation angle of the glass jar through magnetic coupling. A DSLR camera captures projection images from different angles with a laptop PC synchronizing object rotation.
With the captured projection images (usually from 72 angles with 5 degree step size), a 3D root image can be reconstructed with some specific reconstruction codes like Rootwork\cite{Gu2011} or RootReader3D\cite{Clark2014}. Some traits analysis can be conducted in 3D, like root system volume, surface area, total root length, number of branches, etc.
   

\subsection{Imaging protocol}
\begin{figure}
\centering
\includegraphics[width=0.4\textwidth]{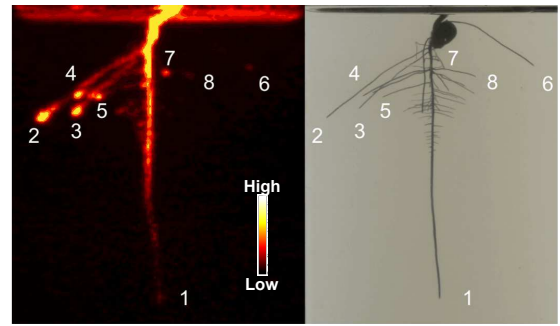}
\caption{ PET (Left) and corresponding optical (Right) images of the maize roots on the 6th day after germination.}
\label{fig:OPT_PET_compare}
\end{figure}

Figure \ref{fig:OPT_PET_compare} shows a young maize plant with structural image acquired from the OPT system and functional image acquired from the plant PET system with $^{11}$CO$_2$ labeling. Those two modalities of root images exhibit the similar root structure, but also indicates some difference, such as the hot spots representing photosynthetic carbon molecules (mainly sucrose) that appear around the root tips in PET image at a later time point (around 111 minutes). Intuitively, these hot spots should be correlated with the most actively growing roots, since the plant must allocate carbon resources to these root tips. With the optical images, the actual root growth rate can be measured.        
A series of studies with a total of 3 subjects have been carried out consecutively with the same imaging protocol to figure out the correlation of root growth rate and activity concentration around root tips. PET scanning started on the day when the first green leaf appeared, and was carried out once a day for 5 days. Each morning, around 10 mCi of $^{11}$CO$_2$ was administrated into a custom built plant labeling chamber. After 6 minutes labeling, residual activity was flushed out with fresh air before a 2$\sim$3 hours PET scan. The same plant was imaged 3 times a day by the OPT system (morning, afternoon and evening with 8 hours apart).

\subsection{Image processing}
\begin{figure}
\centering
\includegraphics[width=0.5\textwidth]{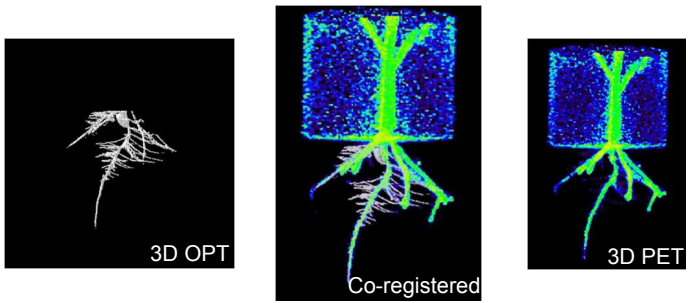}
\caption{ Left: reconstructed 3D optical root image. Right: reconstructed 3D whole plant PET image of a young maize labeled with $^{11}$CO$_2$. Middle: Co-registered 3D PET and OPT images using AMIDE software. Activity concentration represented in PET data is color coded and 3D root images captured from OPT system are in white.}
\label{fig:Coregister_PET_OPT}
\end{figure}

The OPT images are reconstructed using photos from 72 angles. PET images are reconstructed with an ML-EM algorithm\cite{Mathews2013}. The PET and OPT images are aligned using the AMIDE open source software\cite{Loening2003}. Figure \ref{fig:Coregister_PET_OPT} shows the reconstructed 3D OPT and PET images and the co-registered images which are well aligned. Small fine root structures shown in optical image can not be clearly seen in PET image, which partly attributes to the relatively low spatial resolution and partial volume effect in PET imaging\cite{Soret2007a} and is also likely related to biological fact that less carbon is allocated. Some root tips growing close to the wall of glass jar are absent in the optical image, but can be clearly seen in PET images. This relates to the refraction induced distortion in optical images that can not be fully compensated with the rectangular water tank.

The main roots of each subject are selected from optical images and these images are also used to guide the region of interest (ROI) contouring with PET images. The 3D coordinates of the main roots are tracked with different time points and growth rate (mm/day) of the selected roots are calculated based on these time series data points. PET images are first decay corrected and the activity concentration is measured with the ROIs each has a size of 6.4 mm x 6.4 mm x 6.4 mm (8 x 8 x 8 pixels in image).

\section{Results}
\subsection{Dynamic 3D PET images}

\begin{figure}
\centering
\includegraphics[width=0.5\textwidth]{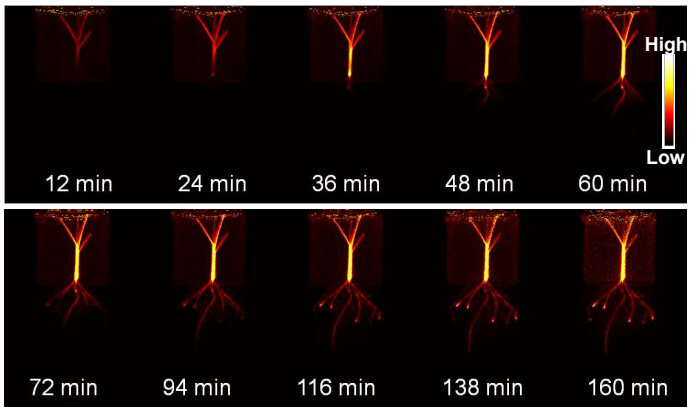}
\caption{ Dynamic PET images with 10 frames acquired with a duration of 160 minutes. The time marked on each frame refers to the start time for acquiring the frame.}
\label{fig:Dynamic_3DPET}
\end{figure}

\begin{figure}
\centering
\includegraphics[width=0.5\textwidth]{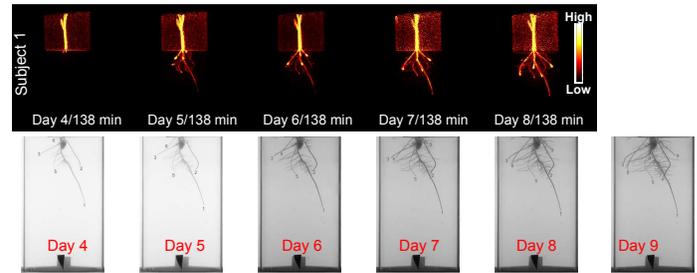}
\caption{ PET images for Subject 1 with last frames of all the five days and the corresponding optical image acquired.}
\label{fig:PEToverDays}
\end{figure}

Each PET image is created with data from 3 different bed positions to cover the entire plant (providing $\sim$28 cm axial FOV). Time presented in Figure \ref{fig:Dynamic_3DPET} and Figure \ref{fig:PEToverDays} are referenced to the beginning of the $^{11}$CO$_2$ injection time. The duration of the first 6 image frames is 12 minutes, which is divided into 0.5 minute for the first bed position to cover the shot part of a plant, 5 minutes for the other two positions to image the stem and root parts and 0.5 minutes for completing the needed mechanical motion. The duration of the last 4 frames is increased to 22 minutes (1 minute, 10 minutes, 10 minutes for the 3 bed positions respectively and 1 minute for mechanical motion) to collect enough events for reconstructing clear PET images. The total duration including the labeling and PET scan reaches 2.5 hours to make sure enough activity is already translocated to the root tips. 

Figure \ref{fig:Dynamic_3DPET} shows the 3D dynamic PET images of Subject 1 acquired on Day 8. Translocation of the $^{11}$C to the root part starts around 30 minutes post-labeling and activity distribution reaches a relatively stable status after two hours. In the last frame of PET images, clear root structure is shown and hot spots appear around the main roots tips. These 5-day PET studies for the 3 subjects show similar dynamic change of $^{11}$CO$_2$ translocation pattern. For these young maize plants, carbon starts to transfer to the root part only after reaching some plant development stage and hot spots appear after that. Figure \ref{fig:PEToverDays} shows the last frame of PET images for Subject 1 of all the 5 days and the actual root growth can be clearly observed for PET images directly.

\subsection{Correlation of activity concentration and roots growth rate}
\begin{figure}
\centering
\includegraphics[width=0.4\textwidth]{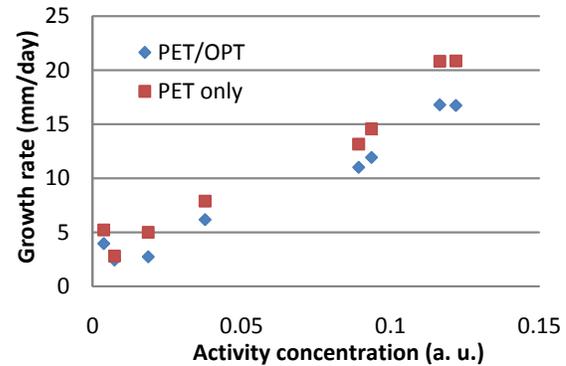}
\caption{ Correlation of roots growth rate with activity concentration around the root tips. The measured data marked with red squares is based on the roots growth rate measured by OPT and activity concentration measured form PET images of selected ROIs. The data shown with blue diamonds is only based on the PET images where the roots growth rate and activity concentration are measured from.}
\label{fig:GrowthRateVsActivity}
\end{figure}

As shown in Figure \ref{fig:PEToverDays}, the root growth rate is measured with 8 selected ROIs of Subject 1 from optical images of Day 7 and Day 8 and the corresponding activity concentration is measured with PET data of Day 7. A good linear correlation is shown in Figure \ref{fig:GrowthRateVsActivity} between activity concentration and root growth rate in this 24 hour window. These data clearly suggest that the activity accumulated at root tips represents carbon allocation by the plant that drives root growth.  

As mentioned above, with these high spatial resolution PET images, the clear root structure appears at later time point of image frames. The 3D coordinates of roots tips can also be measured from the PET images and the roots growth rate can be calculated accordingly. Figure \ref{fig:GrowthRateVsActivity} also shows the the similar linear correlation between activity concentration around roots tips and roots growth rate measured from PET data directly. The result indicates that these kind of studies can be carried out with regular soil using PET only, which may provide more precise data for modeling the relation between carbon allocation and actual root growth.

\subsection{Environmental stress induced root growth rate change}
\begin{figure}
\centering
\includegraphics[width=0.5\textwidth]{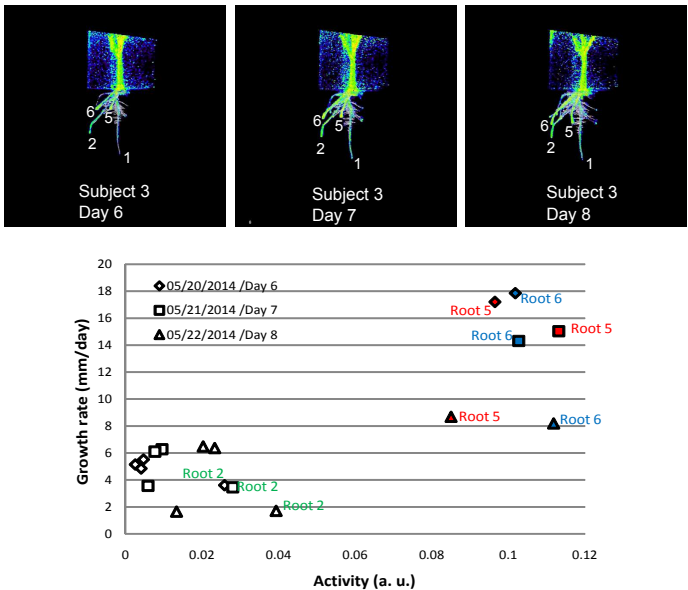}
\caption{Change of correlation between roots growth rate and activity concentration around root tips of Subject 3 measurement from Day 6 to Day 8. Upper: co-registered 3D PET and optical images with different days. Lower: correlation of activity concentration around root tips and roots growth rate of 6 selected main roots measured from Day 6 to Day 8.}
\label{fig:Subject3_GrowthRate_Activtiy}
\end{figure}

\begin{figure}
\centering
\includegraphics[width=0.5\textwidth]{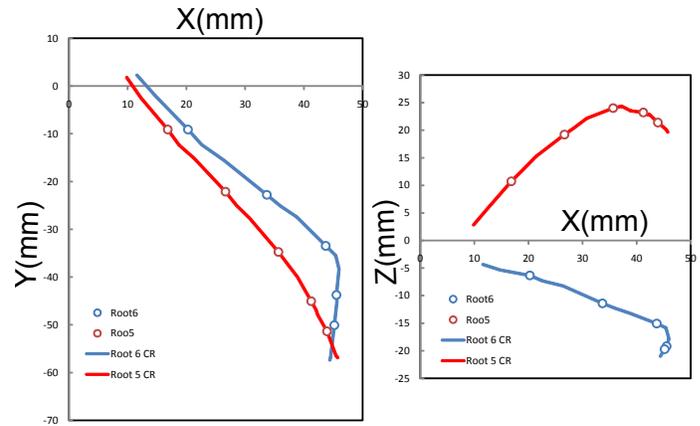}
\caption{Root growth tracks of root 5 and 6 from different projection planes of the Subject 3 maize plant.}
\label{fig:Subject3_root_track}
\end{figure}

Many of the subjects show the similar linear correlation between activity concentration and roots growth rate while some data sets show different and changing relations. Figure \ref{fig:Subject3_GrowthRate_Activtiy} shows the change of correlation between activity concentration at root tips and root growth rate with Subject 3 during the study. The PET data is from Day 6 to Day 8 and optical data is from Day 6 to Day 9. Good correlation still exists between activity concentration at root tips and roots growth rate on Day 6. But on Day 7, the growth rate for root 5 and 6 decreased a lot and a tremendous growth rate decline happened on the following day for those two roots.
The answer to this change can track back to the corresponding optical images. Figure \ref{fig:Subject3_root_track} shows the growth tracks of root 5 and 6 from different projection planes with time stamp marked. The root 5 encountered the wall of the glass jar, and started to change its growth direction and this kind of change happened even a bit earlier for root 6 with the optical image viewed from X-Y plane. These results suggest that the root apical meristem may need to consume more carbon to change its growth direction.

This new observation demonstrates the potential of combined imaging technique in measuring the up/down modulation of molecular processes in plants study when environmental stress arise. 
    
\subsection{Temporal information released with PET image predicates the root growth ahead}
\begin{figure}
\centering
\includegraphics[width=0.5\textwidth]{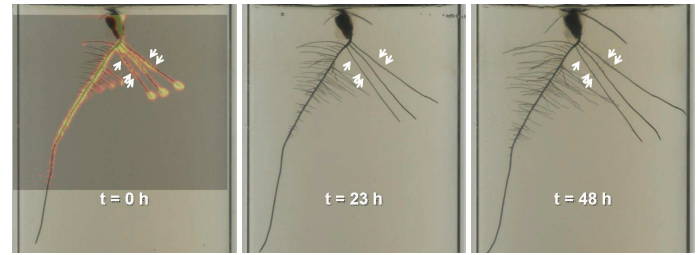}
\caption{Carbon allocation measured by PET (Left) can identify locations of lateral root emergence 48 hours prior to morphological change can be observed (Right).}
\label{fig:Predict_root_growth}
\end{figure}

PET provides near real-time in situ measurement of molecular processes in plants that often precede visible morphological change. This is also observed by our preliminary studies with maize plant. Figure \ref{fig:Predict_root_growth} shows that signatures of carbon allocation can predict lateral root outgrowth around 48 hours prior to the micro-morphological change.

\section{Discussion and Conclusion}
PET provides \textit{in vivo} measurement of dynamic radiotracer distribution in a whole plant non-invasively. Combined PET and optical imaging study of maize roots shows correlation between the activity concentration in the root tips and root growth rate over days. PET also aid in revealing the answer for some plant physiological puzzles by providing 3D dynamic and functional information of a whole plant. 

More applications will be explored by collaborating with plant biologists, combining more imaging modalities, like x-ray CT and hyper-spectral imaging\cite{Furbank2011}\cite{Fiorani2013a}.


%
%
\ifCLASSOPTIONcaptionsoff
  \newpage
\fi



%
\bibliographystyle{IEEEtran}
\bibliography{Combined_PET_OPT_Root_Phenotyping}

\end{document}